\documentclass{ws-p8-50x6-00}

\begin{document}

\title{Recent developments in the dynamical and unitary isobar models for
pion electromagnetic production}

\author{ Shin Nan Yang and Guan Yeu Chen}
\address{Department of Physics, National Taiwan University, Taipei,
Taiwan}
\author{s. S. Kamalov}
\address{Laboratory of Theoretical Physics, JINR, 141980 Dubna,
Russia}
\author{D. Drechsel and  L. Tiator}
\address{Institut f\"ur Kernphysik, Universit\"at Mainz, 55099 Mainz,
Germany}

\maketitle

\abstracts{$\gamma^* N\Delta$ transition form factors and
threshold $\pi^0$ photo- and electroproduction are studied with
the new version of MAID and a dynamical model. By re-analyzing the
recent Jlab data on $p(e,e'p)\pi^0$ at $Q^2 =\, 2.8$ and 4.0
(GeV/c)$^2$, we find that the hadronic helicity conservation is not
yet observed in this region of $Q^2$. The extracted $R_{EM}$,
starting from a small and negative value at the real photon point,
actually exhibits a clear tendency to cross zero and change sign
as $Q^2$ increases, while the absolute value of $R_{SM}$ is
strongly increasing. Our analysis indicates that $A_{1/2}$ and
$S_{1/2}$, but not $A_{3/2} $, starts exhibiting the pQCD scaling
behavior at about $Q^2 \ge 2.5$ (GeV/c)$^2$.  For the $\pi^0$ photo-
and electroproduction near threshold, results obtained within the
dynamical model with the use of a meson-exchange $\pi N$ model for
the final state interaction are in as good agreement with the data
as ChPT.}

\section{Introduction}

Pion photo- and electroproduction reactions are powerful tools to
probe the nucleon structure. The most prominent example is the
$\Delta(1232)$ which decays almost exclusively into $\pi N$
channel. The interest there lies in the possibility of observing
deformation in the $\Delta$. A deformed $\Delta$ would indicate
that the $\Delta$ contains a D-state and the photon can excite a
nucleon through electric $E2$ and Coulomb $C2$ quardrupole
transitions. The study of $E_{1+}^{(3/2)}$ and $S_{1+}^{(3/2)}$
multipoles is expected to shed light on the structure of the
nucleon and its first excited state. Recent
experiments\cite{Beck97} give a nonvanishing ratio
$R_{EM}=E_{1+}^{(3/2)}/M_{1+}^{(3/2)}$ between magnetic dipole
$M_{1+}^{(3/2)}$ and electric quadrupole $E_{1+}^{(3/2)}$
multipoles lying between $ -2.5\%$ and $-3.0\%$ at $Q^2=0$. In
addition, the reaction $p(e,e'p)\pi^0$ provides us with the
possibility  of determining the range of photon four-momentum
transfer squared $Q^2$, where perturbative QCD (pQCD) would become
applicable. In the limit of $Q^2 \rightarrow \infty$, pQCD
predicts the dominance of helicity-conserving
amplitudes\cite{Brodsky81} and scaling results\cite{Carlson}. The
hadronic helicity conservation would have the consequence that
$R_{EM}$ approaches 1. It is an intriguing question how $R_{EM}$
would evolve from a very small negative value at $Q^2 = 0$ to
$+100\%$ at sufficiently high $Q^2$.

Another interesting issue is the $\pi^0$ photo- and
electroproduction in the threshold region. At present time chiral
perturbation theory (ChPT) is the basic theory for the description
of these reactions in this energy region. It predicts a large one
pion loop correction to the low energy theorem (LET) value of the
$E_{0+}$ multipole for the neutral pion
photoproduction\cite{Bernard}. On the other hand the role of the
one-pion loop correction can be related to the effect of final
state interaction (FSI) developed within dynamical
models\cite{Yang1,Nozawa90}. It is of interest to compare the
predictions of these two different approaches.

In this talk, we present the results we have recently obtained
with a new version (hereafter called MAID) of the unitary isobar
model developed at Mainz\cite{MAID98}, and the dynamical model
developed recently in Ref.~\cite{KY99} (hereafter called
Dubna-Mainz-Taipei (DMT) model), both of which give excellent
description of most of the existing pion photo- and
electroproduction data, on the $\gamma^*N\Delta$ transition and
the $\pi^0$ threshold production.

\section{Formalism}

We start with the description of the basic elements of our models.
The main equation for the pion photo- and electroproduction
t-matrix is
\begin{eqnarray}
t_{\gamma\pi}(E)=v_{\gamma\pi}+v_{\gamma\pi}\,g_0(E)\,t_{\pi
N}(E)\,,
\end{eqnarray}
where $t_{\pi N}$ is the full pion-nucleon scattering matrix with
total $\pi N$ c.m. energy $E$ and $g_0(E)$ is the free $\pi N$
propagator.

In resonant channels like (3,3), the transition potential
$v_{\gamma\pi}$ consists of two terms,
$v_{\gamma\pi}(E)=v_{\gamma\pi}^B + v_{\gamma\pi}^R(E)\,,$
where $v_{\gamma\pi}^B$ is the background transition potential and
$v_{\gamma\pi}^R(E)$ corresponds to the contribution of the bare
resonance. The resulting t-matrix can be expressed as a sum of two
terms\cite{KY99},
\begin{eqnarray}
t_{\gamma\pi}=t_{\gamma\pi}^B + t_{\gamma\pi}^R\,,
\label{eq:tgampi}
\end{eqnarray}
where
\begin{eqnarray}
t_{\gamma\pi}^B=v_{\gamma\pi}^B+v_{\gamma\pi}^B\,g_0\,t_{\pi N}\,,
\hspace{1.0cm}
t_{\gamma\pi}^R=v_{\gamma\pi}^R+v_{\gamma\pi}^B\,g_0\,t_{\pi N}\,.
\label{eq:tgampiB}
\end{eqnarray}

For physical multipoles in channel $\alpha=\{l,j\}$, Eq.
(\ref{eq:tgampiB}) gives
\begin{eqnarray}
t_{\alpha}^B(q_E,k;E+i\epsilon)&=&\exp{(i\delta_{\alpha})}\,
\cos{\delta_{\alpha}}\nonumber\\ &\times&\left[v_{\alpha}^B(q_E,k)
+ P\int_0^{~} dq'
\frac{q'^2R_{\alpha}(q_E,q')\,v_{\alpha}^B(q',k)}{E-E(q')}\right],
\label{eq:Tback}
\end{eqnarray}
where $\delta_{\alpha}$ and $R_{\alpha}$ are the $\pi N$
scattering phase shift and reaction matrix, in channel $\alpha$,
respectively; $q_E$ is the pion on-shell momentum and $k=\mid {\bf
k}\mid$ is the photon momentum. The procedure which we use for the
off-shell extrapolation and maintaining the gauge invariance is
given in Ref.~\cite{KY99}.

In the new version of MAID, the $S$, $P$, $D$ and $F$ waves of the
$t_{\alpha}^{B}$ amplitude is defined in accordance with the
K-matrix approximation
\begin{equation}
 t_{\alpha}^{B}({\rm MAID})=
 \exp{(i\delta_{\alpha})}\,\cos{\delta_{\alpha}}
 v_{\alpha}^{B}(q,W,Q^2)\,,
\label{eq:bg00}
\end{equation}
where $W\equiv E$ is the total $\pi N$ c.m. energy. From Eqs.
(\ref{eq:Tback}) and (\ref{eq:bg00}), one finds that the
difference between the background terms of MAID and DMT models is
that pion off-shell rescattering contributions (principal value
integral) are not included in the MAID's background term.

\section{Pion photo- and electroproduction in the $\Delta(1232)$ region}

For the resonance contribution $t^R_\alpha$ in Eq.
(\ref{eq:tgampiB}), the following Breit-Wigner form is
assumed\cite{MAID98} in both models,
\begin{equation}
t_{\alpha}^{R}(W,Q^2)\,=\,{\bar{\cal A}}_{\alpha}^R(Q^2)\,
\frac{f_{\gamma R}(W)\Gamma_R\,M_R\,f_{\pi
R}(W)}{M_R^2-W^2-iM_R\Gamma_R} \,e^{i\phi_R}\,, \label{eq:BW}
\end{equation}
where $f_{\pi R}$ is the usual Breit-Wigner factor describing the
decay of resonance $R$ with total width $\Gamma_{R}(W)$ and
physical mass $M_R$. The phase $\phi_R(W)$ in Eq. (\ref{eq:BW}) is
introduced to adjust the phase of $t_{\alpha}^{R}$ to have the
correct phase $\delta_{\alpha}$.
\begin{figure}[t]
\epsfig{file=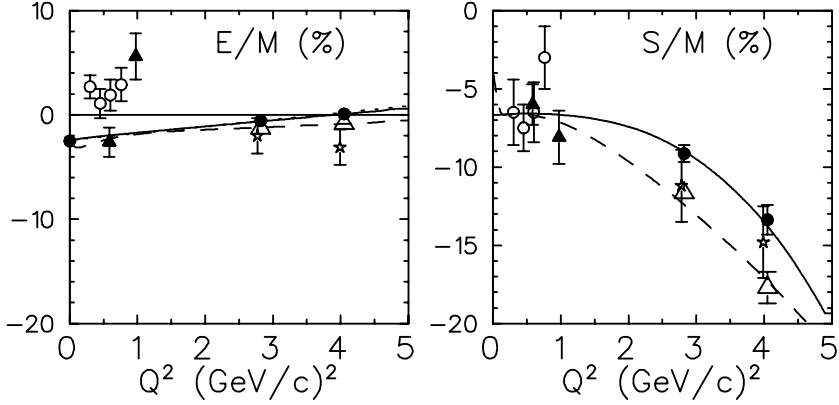,width=11 cm} \caption{ The $Q^2$
dependence of the ratios $R_{EM}^{(p\pi^0)}$ and
$R_{SM}^{(p\pi^0)}$ at $W=1232$ MeV. The solid and dashed curves
are the MAID and DMT models results, respectively, obtained with a
violation of the scaling assumption. Data at $Q^2$ =\,2.8 and 4.0
$(GeV/c)^2$ are from Ref.~\protect\cite{Frolov99} (stars). Results
of our analysis at $Q^2$ =\,2.8 and 4.0 $(GeV/c)^2$ are obtained
using MAID ($\bullet$) and the DMT models ($\bigtriangleup$).
Other data are the same as in Ref.~\protect\cite{KY99}.}
\vspace{-0.5cm}
\end{figure}

The main subject of our study in the resonance region is the
strengths of the electromagnetic transitions described by
amplitudes ${\bar{\cal A}}_{\alpha}^R(Q^2)$. In general, they are
considered as free parameters to be extracted from analysis of the
experimental data. In the present talk we will consider only
results pertinent to the $\Delta(1232)$ resonance. In this case we
impose the following parametrization  for the electric
($\alpha=E$), magnetic ($\alpha=M$) and Coulomb ($\alpha=S$)
amplitudes
\begin{eqnarray}
{\bar{\cal
A}}_{\alpha}^{\Delta}(Q^2)=X_{\alpha}^{\Delta}(Q^2)\,{\bar{\cal
A}}_{\alpha}^{\Delta}(0) \frac{ k}{k_W}\,F(Q^2), \label{eq:Adelta}
\end{eqnarray}
where  $k_W = (W^2 - m_N^2)/2W$. The form factor $F$ is taken to
be $ F(Q^2)=(1+\beta\,Q^2)\,e^{-\gamma Q^2}\,G_D(Q^2)$, where
$G_D(Q^2)=1/(1+Q^2/0.71)^2$ is the usual dipole form factor. The
values of ${\bar{\cal A}}_M^{\Delta}(0)$ and ${\bar{\cal
A}}_E^{\Delta}(0)$ are obtained by setting $X_{\alpha}(0)=1$ and
fitting the pion photoproduction data. The parameters $\beta$ and
$\gamma$ can be determined by setting $X_M^{\Delta}=1$ and fitting
the ${\bar{\cal A}}_{M}^{\Delta}(Q^2)$ to the existing data for
$G_M^*$ form factor.

Note that the physical meaning of the resonant amplitudes in
different models is different\cite{KY99}. In MAID, background
contribution does not contain effects from the off-shell pion
rescattering. Instead, they are effectively included in the
resonance amplitude and it leads to the dressing of ${\bar{\cal
A}}_{\alpha}^{\Delta}(Q^2)$. Thus, using MAID we can extract
information about so the called "dressed" $\gamma N\Delta$ vertex.
However, in the dynamical model the background excitation is
included in $t^{B}_{\alpha}$ and the electromagnetic vertex
${\bar{\cal A}}_{\alpha}^\Delta(Q^2)$ corresponds to the "bare"
vertex.

We determine the $Q^2$ dependence both for $X_E$ and $X_S$,
normalized to 1 at $Q^2=0$, from the recent $p(e,e'p)\pi^0$
experiment\cite{Frolov99} at $Q^2$ = \,2.8 and 4.0 (GeV/c)$^2$.
Note that deviations from $X^\Delta_{\alpha}=1$ value at finite
$Q^2$ will indicate a violation of the scaling law. The extracted
$Q^2$ dependence of the $X_{\alpha}^{\Delta}$ parameters is
$X_{E}^{\Delta}({\rm MAID})=1-Q^2/3.7\,, X_{E}^{\Delta}({\rm
DMT})=1+Q^4/2.4,$ and for the Coulomb $X_{S}^{\Delta}({\rm
MAID})=1+Q^6/61\,, X_{S}^{\Delta}({\rm DMT})=1-Q^2/0.1$. The
obtained  $R_{EM}$ and $R_{SM}=S_{1+}^{(3/2)}/M_{1+}^{(3/2)}$
ratios  are compared with other's results in Fig. 1. The main
difference between our results and those of Ref.~\cite{Frolov99}
is that our values of $R_{EM}$ show a clear tendency to cross zero
and change sign as $Q^2$ increases. This is in contrast with the
results obtained in the original analysis\cite{Frolov99} of the
data which concluded that $R_{EM}$ would stay negative and tends
toward more negative value with increasing $Q^2$. Furthermore, we
find that the absolute value of $R_{SM}$ is strongly increasing.
\begin{figure}[h]
\begin{center}
\epsfig{file=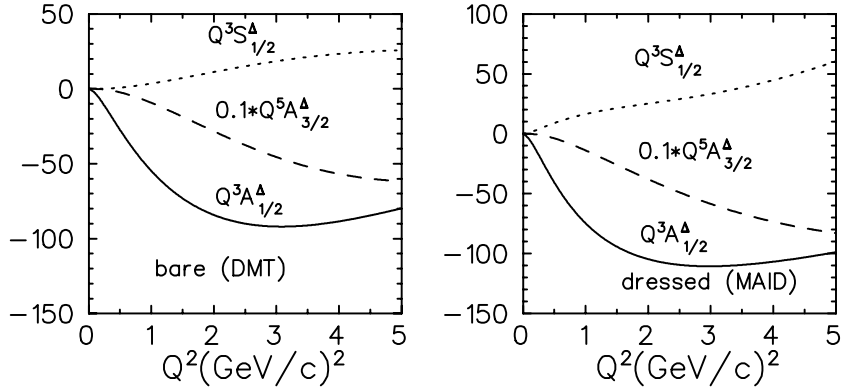,width=11cm}
\end{center}
\caption{Asymptotic pQCD behaviors for the $A_{1/2}$, $A_{3/2}$
and $S_{1/2}$ helicity amplitudes obtained with DMT (left figure)
and MAID (right figure).} \label{fig5}
\vspace{-0.5cm}
\end{figure}

Finally in Fig. 2, we show our results for $Q^3A_{1/2},
Q^5A_{3/2},$ and $Q^3S_{1/2}$ to check the scaling behavior
predicted by pQCD: $A_{1/2}\sim Q^{-3}$, $A_{3/2}\sim Q^{-5}$ and
$S_{1/2}^{(1/2)}\sim Q^{-3}$. It is interesting to see that both
the bare $A_{1/2}$ and $S_{1/2}$ clearly starts exhibiting the
pQCD scaling behavior at about $Q^2 \ge 2.5\, (GeV/c)^2$, while
$A_{3/2}$ does not. From these results, one might be tempted to
speculate that scaling will set in earlier than the helicity
conservation as predicted by pQCD.

\section{Threshold $\pi^0$ photoproduction}

Let us start with the $\pi^0$ photoproduction in the threshold
region. Since the reaction can proceed through the $\pi^0 p$ and $
\pi^+ n$ intermediate states, we may write $\pi^0$ photoproduction
t-matrix as
\begin{eqnarray}
t_{\gamma\pi^0}(E)& = &
v^B_{\gamma\pi^0}+v^B_{\gamma\pi^0}\, g_{\pi^0
p}(E)\,t_{\pi^0 p\rightarrow \pi^0 p}(E)  \nonumber\\ & + &
v^B_{\gamma\pi^+}\, g_{\pi^+ n}(E)\,t_{\pi^+ n\rightarrow \pi^0
p}(E)\,, \label{eq:coupled}
\end{eqnarray}
where $t_{\pi^0 p\rightarrow \pi^0 p}$ and $t_{\pi^+ n\rightarrow
\pi^0 p}$ are the $\pi N$ scattering t-matrices for the elastic
and charge exchange channels, respectively. They are obtained by
solving coupled channels equations for the $\pi N$ scattering
using meson exchange model\cite{MEM}. Results obtained in such an
exact calculation and without the inclusion of FSI are depicted in
Fig. 3 by dash-dotted and dotted curves, respectively. It clearly
indicates that FSI effects is very important and  brings the
results into  agreement with the data. We also find that the main
FSI contribution (around 99\%) comes from the principal value
integral contribution in the charge exchange channel.
\begin{figure}[t]
\epsfig{file=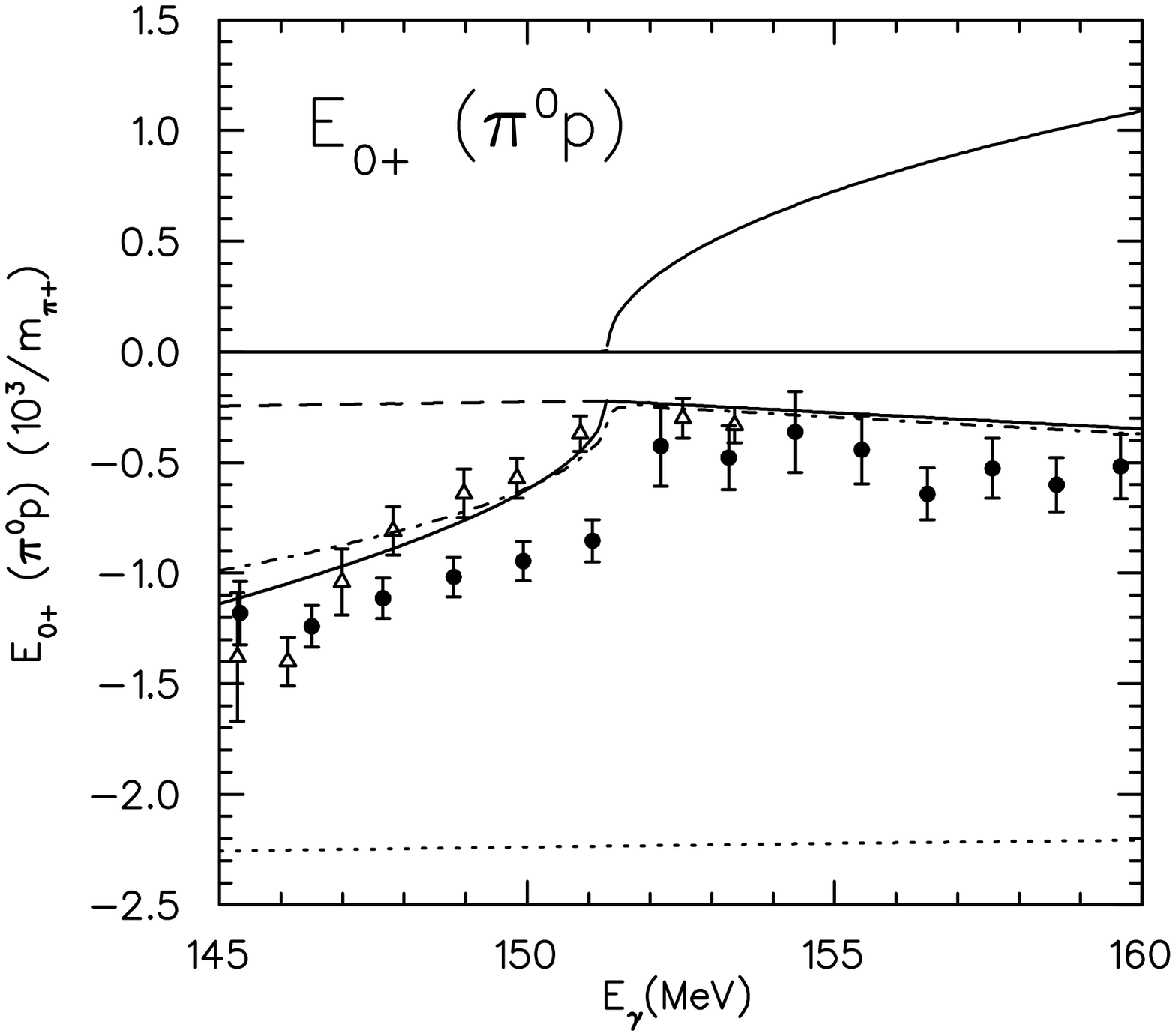, width=5.5 cm}
\epsfig{file=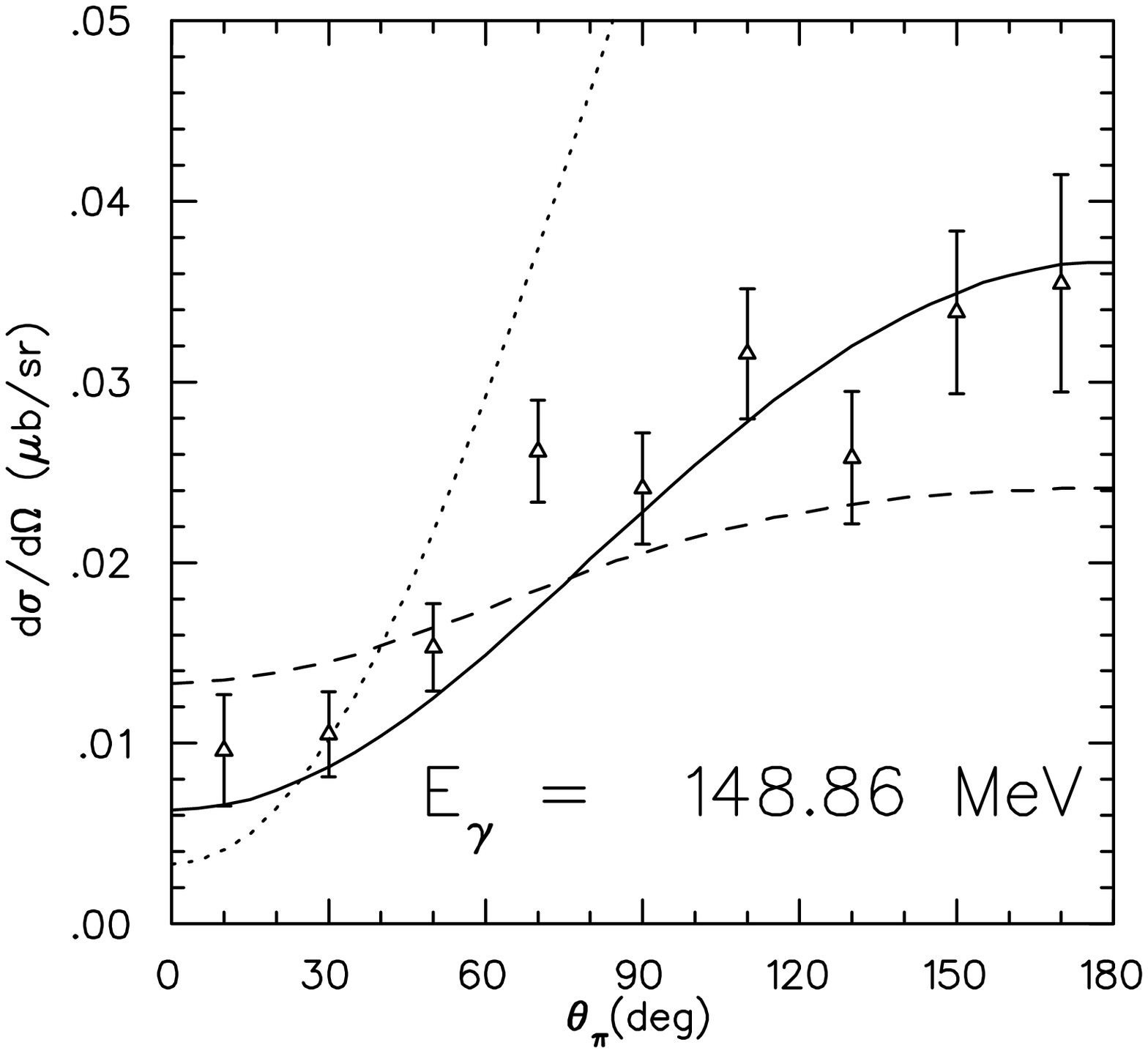, width=5.5 cm} \caption{The $E_{0+}$
multipole (left figure) and differential cross section (right
figure) for the $\pi^0$ photoproduction on the proton. Dotted
curves are the results obtained without FSI. Dash-dotted and solid
curves are the results obtained with coupled-channels and K-matrix
approaches, respectively. Dashed curve is the result without cusp
effect . Data points are from
Ref.~\protect\cite{Fuchs96}($\triangle $) and
Ref.\protect\cite{Bergstrom}($\bullet$)} \vspace{-0.5cm}
\end{figure}

On the other hand, if the FSI effect is evaluated with the
assumption of isospin symmetry (IS), i.e., with averaged masses in
the free $\pi N$ propagator, then the energy dependence in $Re\,
E_{0+}$ near threshold would be smooth as given in the dashed
curve of Fig. 3. Below $\pi^+$ threshold the strong energy
dependence (cusp effect) appears only due to the pion mass
difference and, as mentioned above, it arises mostly from the
coupling with the charge exchange channel. In the literature,
effects from the pion-mass difference below $\pi^+$ production
channel has been taken into account using K-matrix
approach\cite{Bernard,Laget},
\begin{eqnarray}
Re\, E_{0+}^{\gamma\pi^0} = Re\, E_{0+}^{\gamma\pi^0}(IS) - a_{\pi
N} \,\omega_c\,Re
E_{0+}^{\gamma\pi^+}(IS)\,\sqrt{1-\frac{\omega^2}{\omega_c}}\,,
\label{eq:kmatr}
\end{eqnarray}
where $\omega$ and $\omega_c$ are the $\pi^+$ c.m. energies
corresponding to the $W=E_p + E_{\gamma}$ and $W_c=m_n+m_{\pi^+}$,
respectively. $a_{\pi N}=0.124/m_{\pi^+}$ is the pion charge
exchange amplitude, and $E_{0+}^{\gamma\pi^{0,+}}(IS)$ is the
$\pi^0$ and $\pi^+$ photoproduction amplitude at threshold
obtained without pion mass difference using Eq. (\ref{eq:Tback}).
The corresponding results are given by the solid curves in Fig. 3.
We can see that above  $\pi^+$ threshold difference between
results obtained in coupled-channels calculation and K-matrix
approach is very small. This is consistent with the finding of
Ref.~\cite{Laget}.  At lower energies the difference became
visible only very close to the $\pi^0 p$ threshold and it is
around 10\%. In general, we can conclude that Eq. (\ref{eq:kmatr})
is a good approximation for the pion-mass difference effect. The
threshold energy dependence for imaginary part can be easily
obtained via Fermi-Watson theorem if  in the threshold region $\pi
N$ phase shift is taken as a linear function of the $\pi^+$
momentum, i.e. they approaches 0 at $q_{\pi^+}\rightarrow 0$.

In Fig. 3, we also compare our prediction for the differential
cross section with the data from Mainz\cite{Fuchs96}. We see that
both the off-shell pion rescattering and cusp effects
substantially improve agreement with the data. Moreover it
indicates that our model, without any new free parameter, also
gives reliable predictions for the $p$-waves threshold behaviour.

\section{Threshold $\pi^0$ electroproduction}

Pion electroproduction provides us with information on the $Q^2$
dependence of the transverse $E_{0+}$ and longitudinal $L_{0+}$
multipoles in the threshold region. The "cusp" effects in $L_{0+}$
multipole is taken into account in a similar way as in the case of
$E_{0+}$,
\begin{eqnarray}
Re\, L_{0+}^{\gamma\pi^0} = Re\, L_{0+}^{\gamma\pi^0}(IS) - a_{\pi
N} \,\omega_c\,Re\,
L_{0+}^{\gamma\pi^+}(IS)\,\sqrt{1-\frac{\omega^2}{\omega_c}}\,.
\end{eqnarray}
At threshold, the $Q^2$ dependence is given mainly by the Born +
vector mesons contributions in $v_{\gamma\pi}^B$, as described in
Ref.~\cite{MAID98}.
\begin{figure}[t]
\epsfig{file=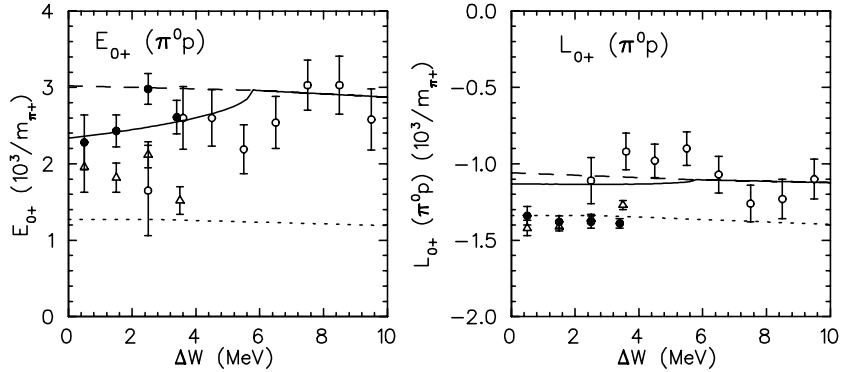, width=11 cm} \caption{ Real parts of the
$E_{0+}$ (left figure) and $L_{0+}$ (right figure) multipoles for
the $\pi^0$ electroproduction on the proton at $Q^2$=0.1
(GeV/c)$^2$. Notations for the curves are the same as in Fig. 1.
Data points are from Ref.\protect\cite{NIKHEF}($\circ$) and
Ref.\protect\cite{Distler}($\triangle $). Results of the present
work obtained using $p$-waves from the DMT model are given by
($\bullet$).} \vspace{-0.5cm}
\end{figure}
In Fig. 4 we show our results for the cusp and FSI effects in
$E_{0+}$ and $L_{0+}$ multipoles for the $\pi^0$ electroproduction
at $Q^2=0.1$ (GeV/c)$^2$, along with the results of the multipole
analysis from NIKHEF\cite{NIKHEF} and Mainz\cite{Distler}. Note
that results of both groups were obtained using the $p$-wave
predictions given by ChPT. We have made a new analysis of the
Mainz data\cite{Distler} for the differential cross sections,
using our DMT $p$-wave multipoles instead. The $s$-wave multipoles
extracted this way  are also shown in Fig. 2 (solid circles). Note
that results of our new analysis for the $E_{0+}$ multipole are
closer to the NIKHEF data and in better agreement with DMT model
prediction. However, the results for the longitudinal $L_{0+}$
multipole stay practically the same as in the previous Mainz
analysis. We see that in this case DMT model prediction is in
better agreement with the NIKHEF data.

In Fig. 5, DMT model predictions (dashed curves) are compared with
the Mainz experimental data  for the unpolarized cross sections
$d\sigma/d\Omega=d\sigma_T/d\Omega + \epsilon\,d\sigma_L/d\Omega$,
and longitudinal-transverse cross section $d\sigma_{TL}/d\Omega$.
Overall, the agreement is good. If the $L_{0+}$ multipole in DMT
model is replaced with that extracted from Mainz data, then the
agreement with the Mainz data is further improved as given by the
solid curves in Fig. 5.
\begin{figure}[h]
\epsfig{file=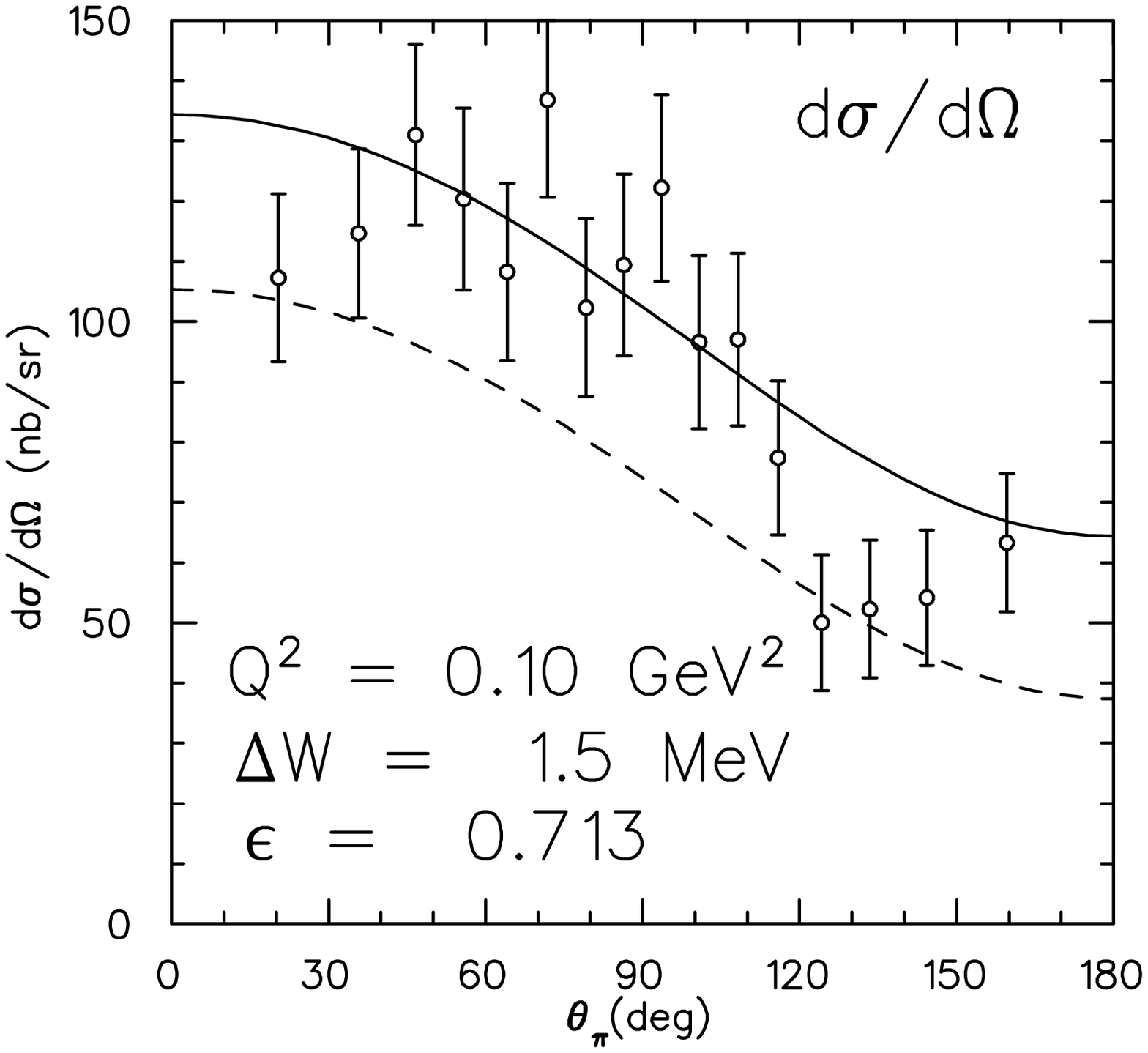, width=5.5 cm}
\epsfig{file=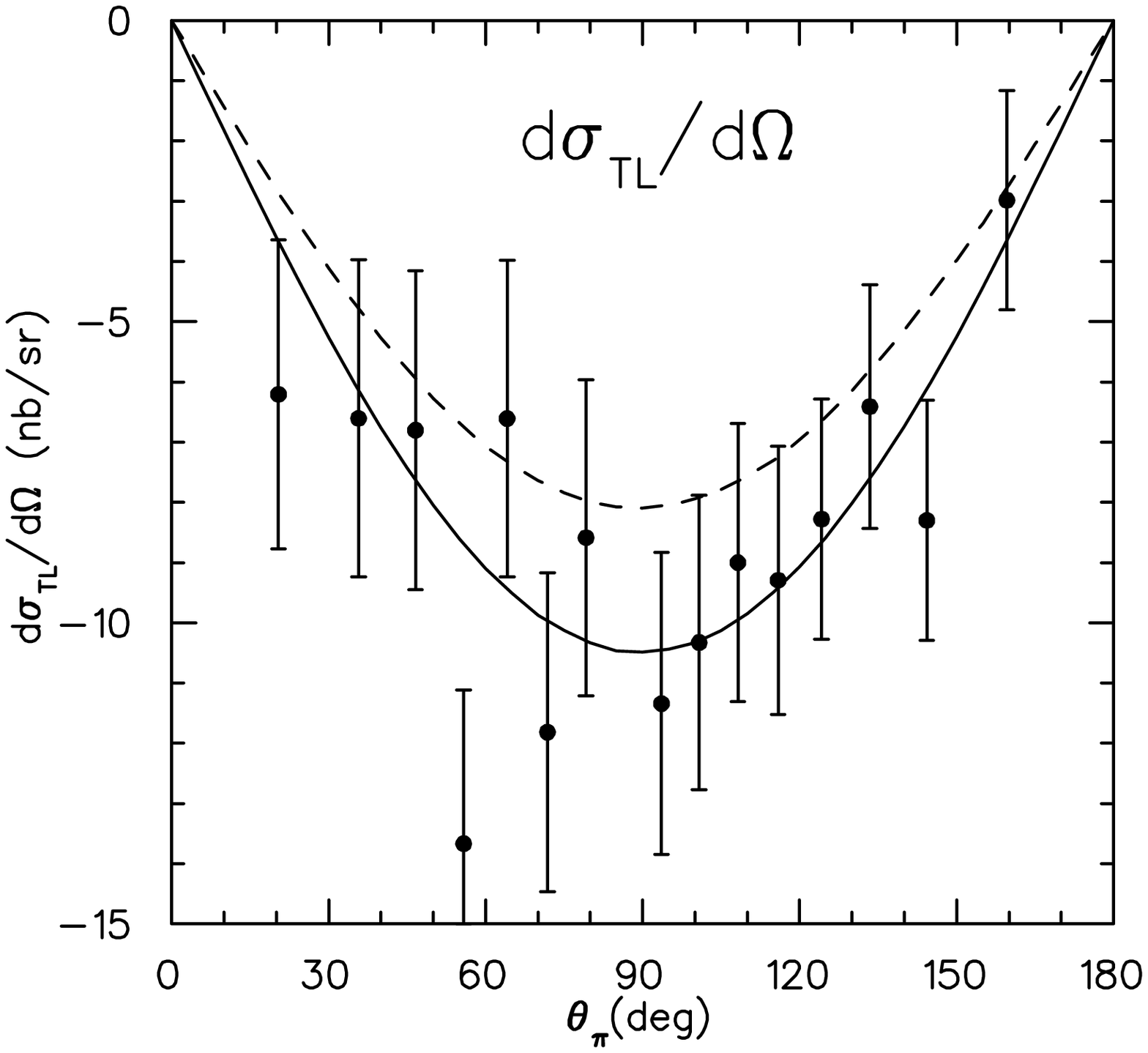, width=5.5 cm} \caption{ Angular
 distribution $d\sigma/d\Omega=d\sigma_T/d\Omega +
\epsilon\,\sigma_L/d\Omega$, for the $\pi^0$ electroproduction on
the proton at $Q^2$=0.1 (GeV/c)$^2$ , $\epsilon=0.713$ and at
total c.m. energies  $\Delta W=W-W_{thr}^{\pi^0p}=1.5$ MeV. Dashed
curves are the predictions of the DMT model. Solid curves are our
results of the local fit with fixed $p$-waves. Experimental data
from Ref.\protect\cite{Distler}.} \vspace{-0.5cm}
\end{figure}

\section{Conclusion}

In summary, the $\gamma^* N\Delta$ transition form factors and
threshold $\pi^0$ photo- and electroproduction are studied with
the new version of MAID and a dynamical model. By re-analyzing
recent Jlab data on $p(e,e'p)\pi^0$ at $Q^2 =\, 2.8$ and $4.0\,
(GeV/c)^2$, we find that $A_{3/2}$ is still as large as $A_{1/2}$
at $Q^2=4$ (GeV/c)$^2$, which implies that hadronic helicity
conservation is not yet observed in this region of $Q^2$.
Accordingly, our extracted values for $R_{EM}$ are still far from
the pQCD predicted value of $+100\%$. However, in contrast to
previous results we find that $R_{EM}$, starting from a small and
negative value at the real photon point, actually exhibits a clear
tendency to cross zero and change sign as $Q^2$ increases, while
the absolute value of $R_{SM}$ is strongly increasing. In regard
to the scaling, our analysis indicates that $A_{1/2}$ and
$S_{1/2}$, but not $A_{3/2}$, starts exhibiting the pQCD scaling
behavior at about $Q^2 \ge 2.5 (GeV/c)^2$. It appears likely that
the onset of scaling behavior might take place at a lower momentum
transfer than that of hadron helicity conservation. For the
$\pi^0$ photo- and electroproduction near threshold, results
obtained within the dynamical model with the use of a
meson-exchange  $\pi N$ model for the final state interaction are
in as good agreement with the data as ChPT.

\end{document}